\documentclass[%
 reprint,
 amsmath,amssymb,
 aps,
]{revtex4-2}

\usepackage{graphicx}
\usepackage{dcolumn}
\usepackage{color}
\usepackage{CJK}
\usepackage{fancyhdr}
\usepackage{calc}
\usepackage{lineno}
\usepackage{float}
\usepackage{mpgmpar} 
\usepackage{amsmath,amsthm} 
\usepackage{bm,here}
\usepackage{lipsum}
\usepackage{ragged2e}
\usepackage{booktabs}
\usepackage{braket, physics}
\usepackage{ulem}

\begin{document}
\preprint{APS/123-QED}

\title{Transient Pauli blocking in a InN film as a mechanism for broadband ultrafast optical switching}
\author{Junjun Jia}
\email{jia@aoni.waseda.jp}
\affiliation{Global Center for Science and Engineering (GCSE), Faculty of Science and Engineering, Waseda University, 3--4--1 Okubo, Shinjuku, Tokyo 169--8555, Japan.}%
\affiliation{Graduate School of Advanced Science and Engineering, Waseda University, 3--4--1 Okubo, Shinjuku, Tokyo 169--8555, Japan}%

\author{Minseok Kim}
\affiliation{Graduate School of Science and Engineering, \\Aoyama Gakuin University, Sagamihara, Kanagawa, 252--5258, Japan.
}%

\author{Yuzo Shigesato}
\affiliation{Graduate School of Science and Engineering, \\Aoyama Gakuin University, Sagamihara, Kanagawa, 252--5258, Japan.
}%


\author{Ryotaro Nakazawa}
\affiliation{Institute for Molecular Science, 38, Nishigounaka, Myodaiji, Okazaki, Aichi, 444--8585, Japan}%

\author{Keisuke Fukutani}
\affiliation{Institute for Molecular Science, 38, Nishigounaka, Myodaiji, Okazaki, Aichi, 444--8585, Japan.}%
\affiliation{The Graduate University for Advanced Studies (SOKENDAI), Hayama--cho, Miura--gun, Kanagawa 240--0193, Japan.}

\author{Satoshi Kera}
\affiliation{Institute for Molecular Science, 38, Nishigounaka, Myodaiji, Okazaki, Aichi, 444--8585, Japan.}%
\affiliation{The Graduate University for Advanced Studies (SOKENDAI), Hayama--cho, Miura--gun, Kanagawa 240--0193, Japan.}

\author{Toshiki Makimoto}
\affiliation{ Graduate School of Advanced Science and Engineering, Waseda University, 3--4--1 Okubo, Shinjuku, Tokyo 169--8555, Japan}%

\author{Takashi Yagi}
\affiliation{%
   National Metrology Institute of Japan (NMIJ), 
   \\National Institute of Advanced Industrial Science and Technology (AIST), 
   \\Central 5, 1--1--1 Higashi, Tsukuba, Ibaraki 305--8565, Japan
}%

\date{\today}

\begin{abstract}
The transient Pauli blocking effect offers a promising route for achieving ultrafast optical switching in semiconductors, enabling a rapid switching from an initially opaque state to a relatively transparent state upon photoexcitation. Herein, we demonstrate broadband ultrafast optical switching in degenerate InN thin films, spanning the visible to near--infrared spectral range, using pump--probe transient transmittance measurements. To elucidate the underlying physical mechanism, we perform probe--energy--resolved analysis for ultrafast dynamics, and develop a theoretical model based on a quasi--equilibrium Fermi--Dirac distribution. The model successfully captures the experimental transients and yields an electron--phonon coupling constant of 1.0 $\times$ 10$^{17}$ W/m$^3$$\cdot$K, along with an electronic specific heat coefficient ranging from 1.52 to 2.02 mJ/mol$\cdot$K$^2$, which allow direct prediction of the spectral switching window. Notably, we demonstrate that the Pauli blocking effect can be induced solely by a laser--excitation driven rise in electronic temperature, without requiring significant carrier injection into the conduction band in degenerate semiconductors. These findings offer new insights for designing ultrafast optical modulators, shutters, and photonic devices for next-generation communication and computing technologies.

\end{abstract}

\maketitle

\section{Introduction}
With the rapid advancement of high-intensity laser technology, integrating laser irradiation with materials, which can generate various nonequilibrium excitation states, has opened up new avenues for the design of functional materials and devices \cite{Jia2022, Jia2025}. Since most commercial lasers operate at wavelengths extending upto the ultraviolet region, semiconductor materials have become ideal platforms for generating light-driven functionalities, such as optical switching devices \cite{Schlaepfer2018}, second-- \cite{Lafrentz2013} or higher-- order harmonic generation \cite{Liu2017, Freeman2022}, thereby enabling a wide range of emerging applications.

\begin{figure}[ht]
\begin{center}
\includegraphics[clip, width=8cm]{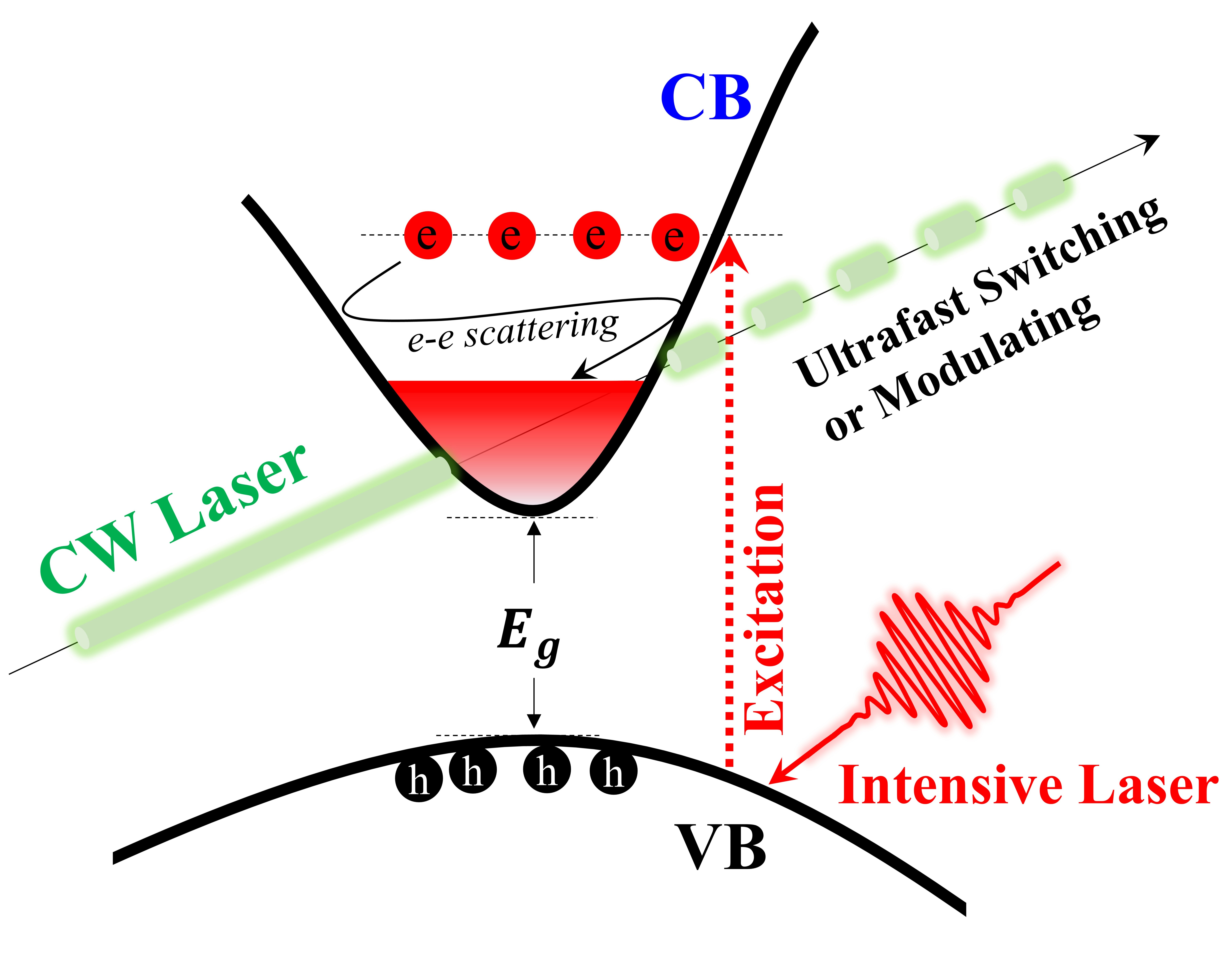}
\caption{\textbf{Schematic illustration of ultrafast optical switching in semiconductors via transient Pauli blocking.} Under intensive laser irradiation with photon energy exceeding the band gap $E_g$, electrons are excited into conduction band and rapidly thermalize through electron--electron scattering ($e-e$ scattering), forming a transient quasi--equilibrium Fermi-Dirac distribution. The resulting electronic occupation  transiently blocks optical absorption, enabling an ultrafast switching from an opaque to a transparent state. This mechanism offers opportunities for laser pulse modulation and other advanced optical applications. CB: conduction band, VB: valence band.}
\label{schematic}
\end{center}
\end{figure}

When semiconductors are irradiated with intense pulsed laser with photon energy exceeding the bandgap, a large number of electrons are excited from the valence band into the conduction band. These photoexcited electrons rapidly undergo energy and momentum relaxation owing to the fast electron--electron interaction, thermalizing from their initial non--thermal distribution to a quasi--equilibrium Fermi--Dirac (FD) distribution characterized by an elevated electronic temperature and then cools down mainly by electron-phonon scattering \cite{Allen1987, Crepaldi2012}, as illustrated in Fig. \ref{schematic}, which subsequently cools mainly via carrier--phonon scattering. Before recombining with holes left in the valence band, the thermalized electrons transiently occupy the conduction band on ultrafast timescales, ranging from femtoseconds to nanoseconds. This electronic occupation gives rise to a \textit{transient Pauli blocking effect}, which blocks absorption of photons with energies at or slightly above the bandgap \cite{Jia2025}. As a result, the material undergoes an ultrafast switching from an initial opaque to a transiently transparent state, significantly modulating its optical properties of the semiconductor on ultrafast timescales. Such an effect offers new opportunities to exploit the transient state as a novel functional phase for optical applications, including ultrafast optical switches, filters and modulator, as shown in Fig. \ref{schematic}. 

How should we interpret the influence of \textit{transient Pauli blocking effect} on the optical properties of semiconductors? Once the conduction band is populated with a thermalized electron gas, a notable modification to the optical response is expected from collective motion of these carriers. This behavior can be treated within the framework of classical Drude model. According to the Drude model, light with frequencies below the plasma frequency is reflected, while those with higher frequencies can propagate through the material \cite{Jia2014, Jia2020}. Photoexcitation--induced changes in carrier density can thus modulate the plasma frequency and, in turn, the optical properties of the material. However, even at high excited carrier density upto 10$^{21}$ cm$^{-3}$, the collective response remains relatively weak in the visible spectrum, as those observed in heavily doped In$_2$O$_3$ films \cite{Jia2019}. Therefore, this collective behavior is expected to manifest primarily in the near--infrared region and long--wavelength regions. 

On the other hand, the transient occupation of thermalized electrons in the conduction band often leads to a temporary increase in transmittance in the visible region \cite{Jia2022, Pacebutas2006}, a phenomenon not adequately explained by the conventional Drude model. Due to its ultrafast and nonequilibrium nature, this behavior lies beyond the scope of classical optical transition theories developed under equilibrium assumptions \cite{Tiwari2024}. Although advanced theoretical approaches, such as time-dependent density functional theory (TDDFT) \cite{Sato2014, Schultze2014} are being actively explored in this field, a unified theory that incorporates direct/indirect optical transitions \cite{Tiwari2024} with band gap renormalization \cite{Jia2013} and the collective response of thermalized electrons and so on is still lacking. Moreover, the substantial computational cost of these methods hinders their practical use in interpreting experimental results from ultrafast spectroscopy. From an optical perspective, the transient occupation of thermalized electrons in the conduction band blocks previously allowed optical transitions, thereby reducing absorption and giving rise to a transient increase in transmittance. However, the physical origin of the \textit{transient Pauli blocking effect} and its manifestation as a broad spectral feature in the visible range remains insufficiently understood \cite{Chen2003, Sun2008}, highlighting the need for a more simplified theoretical framework. 

In this study, we investigate the \textit{transient Pauli blocking effect} in a degenerate wurtzite--structured InN film using pump–-probe transient transmittance measurements with multicolor probe lasers, different from our previously reported degenerate pump-–probe approaches \cite{Jia2022}. We develop a theoretical model that captures the observed ultrafast electronic dynamics, and reveals a new physical origin of the \textit{transient Pauli blocking effect}. Specifically, we find that \textit{transient Pauli blocking effect} can be induced solely by a laser–excitation driven rise in electronic temperature, without requiring significant carrier injection into the conduction band. Moreover, our model defines the spectral window over which optical switching occurs. This framework not only deepens fundamental understanding of nonequilibrium carrier dynamics including its measurement, but also opens new avenues for realizing semiconductor--based ultrafast optical switches, filter and modulators leveraging \textit{transient Pauli blocking effect} as an active control mechanism.

\section{Experimental and computational details}
Molecular epitaxial evaporation was used to grow the $c$--oriented InN film on a $c$--sapphire substrate. The fabricated InN sample is polycrystalline $n$ type, with an electron concentration of 5.50 $\pm$ 0.82$\times$10$^{19}$ cm$^{-3}$ and a Hall mobility of 5.43 $\pm$ 0.64 cm$^{2}$/Vs using a HL--5500PC Hall measurement system (Bio-Rad). The film thickness was determined to be 400 nm by fitting the Fourier--transform infrared (FTIR) spectrum using the Drude model. The out--of--plane lattice constants $c$, calculated from the X--ray diffraction (XRD) (002) peak at 2$\theta$=31.36172$^\circ$, was found to be 5.700 \AA, in good agreement with the standard powder diffraction value of 5.7033~\text{\AA} at room temperature \cite{Paszkowicz1999, Kuzmik2017}. The absorption coefficient $\alpha$ was extracted from optical transmittance ($T$) and reflectance ($R$) spectra measured using a UV--Vis--NIR spectrometer (UV--3600 plus, Shimadzu, Japan), based on the relation $T=(1-R)\exp(-\alpha d)$, where $d$ is the film thickness. 

High--sensitivity ultraviolet photoelectron spectroscopy (HS-UPS) was performed to determine the electronic structure around the top of valence band for the used InN film at the Institute for Molecular Science. The base pressure during the measurements was approximately 10$^{-8}$ Pa. A bias voltage of –5 V was applied to the sample. A monochromatic Xe I$\alpha$ line (photon energy: 8.437 eV) was used as the excitation source. The light incident angle was 45$^\circ$ with respect to the sample surface. The HS-UPS measurements were performed using the hemispherical analyzer (MBS A-1) in the normal emission geometry, where the sample’s surface-normal was aligned along the analyzer axis, and the pass energy was set to 10 eV. The total energy resolution was set to 75 meV. The photoelectrons’ kinetic energy at the Fermi level and the instrumental energy resolution were calibrated using the Fermi edge of a gold film measured at room temperature.

Femtosecond time--resolved transient transmission measurements were conducted using a pump--probe technique to investigate ultrafast carrier dynamics. The InN film was excited at room temperature by a 140 fs Ti:Sapphire laser pulse (Repetition rate: 76 MHz, photon energy: 1.55 eV). The laser beam was split using a polarized beam splitter: the majority of the power was directed to the sample as the pump, while the remaining power was coupled into a supercontinuum generation module (FemtoWHITE 800, NKT Photonics) to produce the broadband probe laser spanning 400--1000 nm. Both pump and probe beams were focused normal to the sample surface (incident angle: 90$^\circ$) using a 10$\times$ objective lens (Mitutoyo). The pump beam spot diameter (1/$e^2$ diameter) was 10.3 $\mu$m as determined by the knife--edge method. The transmitted probe light was spectrally resolved using a grating spectrometer and subsequently detected by a silicon photodetector (2107--FC, New Focus). 

To interpret the ultrafast dynamics observed in the transient transmission measurements, the photoexcited carrier density ($n_{ex}$) was estimated using the method proposed by J. Menéndez \textit{et al.} \cite{Menendez2020}, which accounts for the lateral diffusion of excited electrons during pumping. In this estimation, the room--temperature average carrier lifetime was conservatively set to 5 ps based on our pump--probe measurements, the carrier diffusion coefficient was taken as 0.14 cm$^2$/s derived from the measured Hall moblity, and the surface recombination velocity was assumed to be 2 $\times$ 10$^7$ cm/s. For a pumping power of 220 mW at 1.55 eV, $n_{ex}$ was estimated to be 3.84 $\times$ 10$^{16}$ cm$^{-3}$, for which the measured $R$=0.136 and $\alpha$=29346 cm$^{-1}$ were used.

Density functional theory (DFT) calculations were performed using the Vienna \textit{ab initio} Simulation Package \cite{Kresse1996} with the projector augmented--wave method \cite{Kresse1999}. The experimentally measured lattice constants were used, while keeping the fractional coordinates fixed, to calculate the band structure of wurtzite InN (space group No. 186, $P6_3mc$) using the Heyd--Scuseria--Ernzerhof (HSE06) hybrid exchange correlation functional with a plane--wave cutoff of 400 eV. The spin--orbit coupling effect was included to accurately capture the valence band splitting. Optical transition matrix elements were calculated in the longitudinal gauge \cite{Rödl2019}. They are given as matrix elements $\matrixel{c\mathbf{k}}{\mathbf{p}}{v\mathbf{k}}$  of the momentum operator $\mathbf{p}$ between the conduction band $c$ and the valence band $v$ at a given $\mathbf{k}$ point. Considering the spin--orbit degenerte states $i,j$=1, 2 in the conduction and valence bands, the magnitude of matrix elements for direct transitions can be expressed as 

\begin{equation}
  |\mathbf{M_{cv}}|=\sum_{c_i, v_j}\matrixel{c_i\mathbf{k}}{\mathbf{p}}{v_j\mathbf{k}}.
\end{equation}

Then, the transition probability can be proportional to the optical matrix element, namely $p_{cv} \propto \mathbf{|M_{cv}|}^2$.

\section{Results and Discussion}

\begin{figure}[ht]
\begin{center}
\includegraphics[clip, width=8cm]{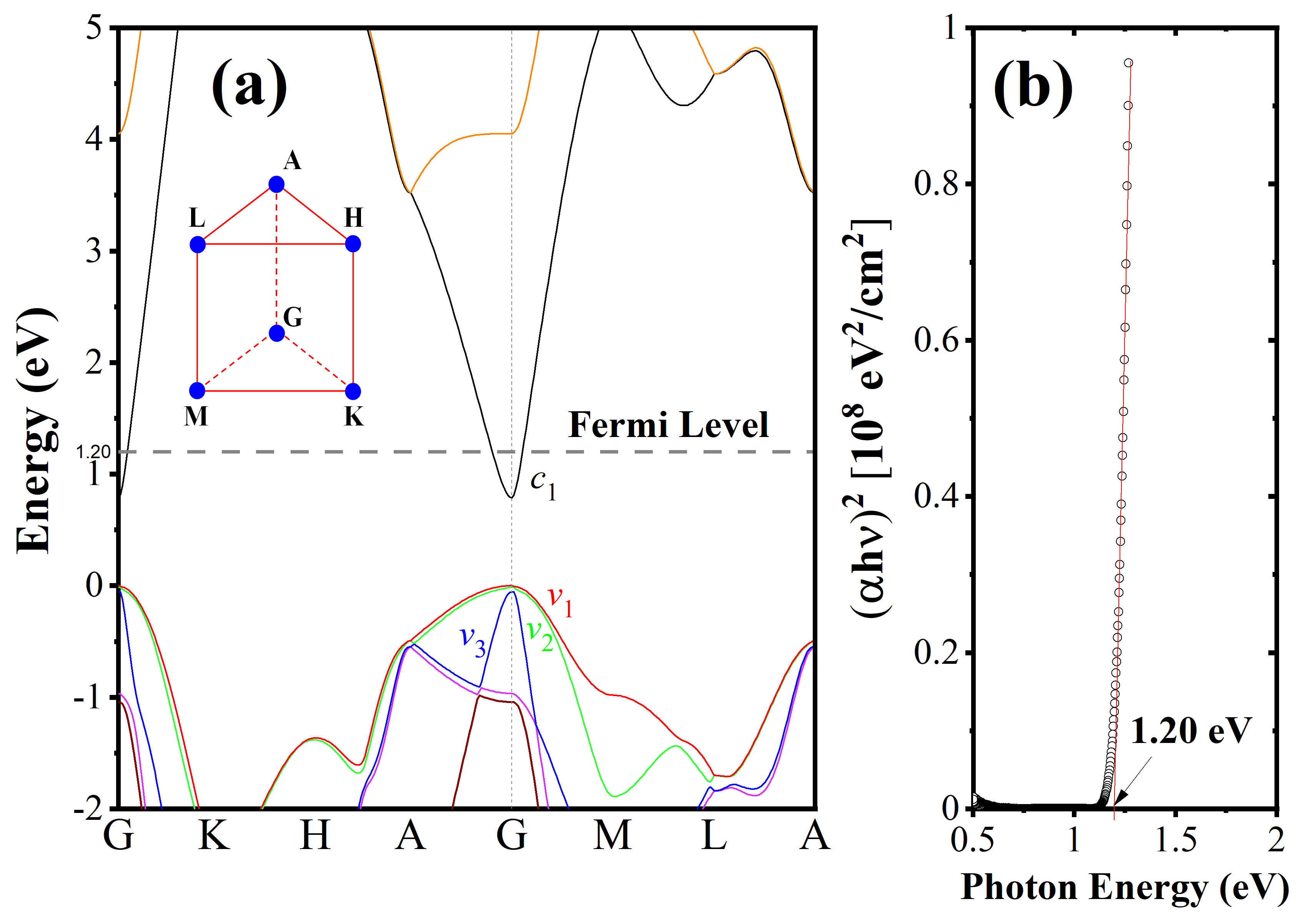}
\caption{\textbf{Electronic band structure and optical band gap of wurtzite InN.} (a) Calculated band structure of wurtzite InN using the HSE hybrid functional with spin--orbit coupling interaction included. The inset shows the high--symmetry points in the first Brillouin zone. The dashed line represents the Fermi level, estimated based on the unintentional carrier density obtained from Hall effect measurements. (b) Tauc plot used to extract the optical band gap from the experimental data (shown as circles).}
\label{HSE}
\end{center}
\end{figure}

To elucidate the \textit{transient Pauli blocking effect} in InN within the frame of band theory, we started by calculating the band structure of wurtzite--structured InN using the HSE06 functional, incorporating spin--orbit coupling effect to account for valence band splitting. The calculation was performed using experimentally determined lattice constants ($a$=3.533~\AA\ and $c$=5.700~\AA) at room temperature. The calculated band structure is shown in Fig. \ref{HSE}(a), revealing a band gap of 0.79 eV at the $\Gamma$ point. This value is in good agreement with previous theoretical calculations \cite{Furthmuller2005} and the experimentally measured band gap of $\sim$0.80 eV obtained from PL measurements on low--carrier--concentration InN sample \cite{Davydov2002}. Given its degenerate nature of the used InN sample, arising from its high electron concentration, the position of Fermi level was estimated based on the Burstein--Moss effect \cite{BM}. The corresponding Fermi level shift ($ \Delta E$) is calculated by 
\begin{equation}
 \Delta E=\frac{\hbar^2}{2m_e^*}(3\pi^2 n_0)^{\frac{2}{3}}, 
\label{bandfill}
\end{equation}
where $\hbar$ is the reduced Planck constant, $m_e^*$ is the electron effective mass, and $n_0$ is the electron concentration. Because the $\bf k \cdot \bf p$ interaction across the narrow direct gap of InN causes strong nonparabolic $E$--$k$ dispersion in the CB \cite{Kane1957, Wu2003}, an increased $m_e^*$ away from the CB minimum must be taken into account \cite{Jia2022, Wu2003, Fu2004}. For a carrier density of 5.50 $\times$ 10$^{19}$ cm$^{-3}$, $m_e^*$ is estimated to be 0.13$m_0$, where $m_0$ is the free electron mass, resulting in a Fermi level shift of approximately 0.40 eV. Consequently, the effective optical band gap is estimated to be 1.19 eV. This value is in close agreement with the optical band gap extracted from Tauc's plot in Fig. \ref{HSE}(b).

\begin{figure}[ht]
\begin{center}
\includegraphics[clip, width=8cm]{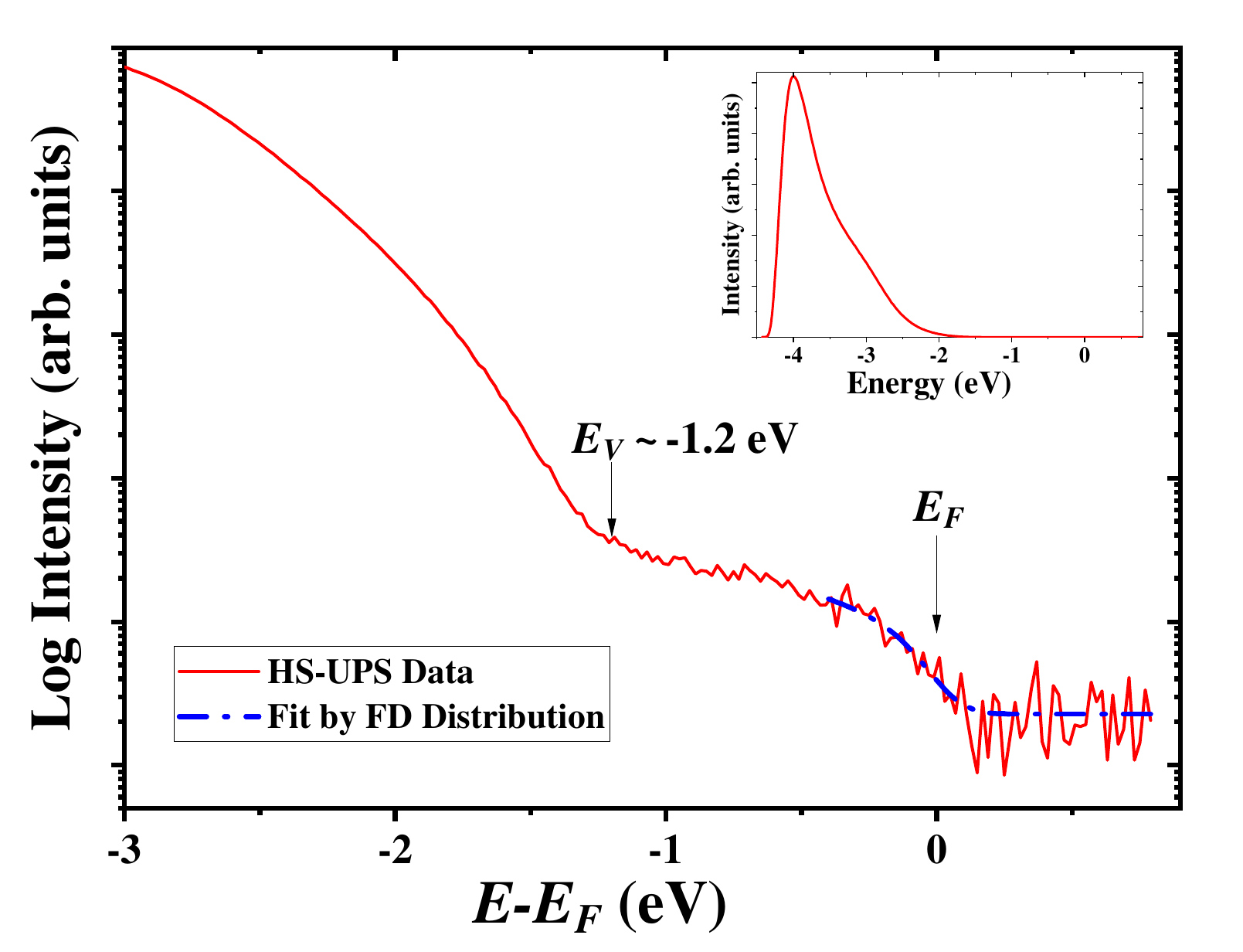}
\caption{\textbf{HS--UPS spectrum of the InN thin film.} HS--UPS spectrum obtained using the monochromatic Xe I$\alpha$ line (photon energy of 8.437 eV), plotted on a logarithmic scale. The dash--dotted line near 0 eV represents the fitting curve based on the FD distribution. The inset shows the same spectrum on a linear scale. The labels ``$E$", ``$E_F$", and ``$E_V$" indicate the energy, Fermi level, and the valence band maximum, respectively.}
\label{XPS}
\end{center}
\end{figure}

To further determine the Fermi level, the HS--UPS spectrum of the InN thin film was collected, as shown in Fig. \ref{XPS}. The horizontal axis indicates the energy relative to the Fermi level, and the vertical axis represents the intensity of the photoelectrons. On a logarithmic scale, a weak but continuous signal can be observed up to 0 eV. A change in the spectral slope appears around –1.2 eV, which is assigned to the valence band maximum (VBM). The electronic states observed between the VBM and the Fermi level arise from in--gap states and electrons occupying the conduction band. The line shape near 0 eV is well reproduced by a FD distribution, indicating partial occupation of the conduction band. Because the InN film is degenerate, the energy difference between the Fermi level and the VBM corresponds approximately corresponds to the optical band gap, consistent with the value obtained from the Tauc plot in Fig. 2(b).

\begin{figure}[ht]
\begin{center}
\includegraphics[clip, width=9cm]{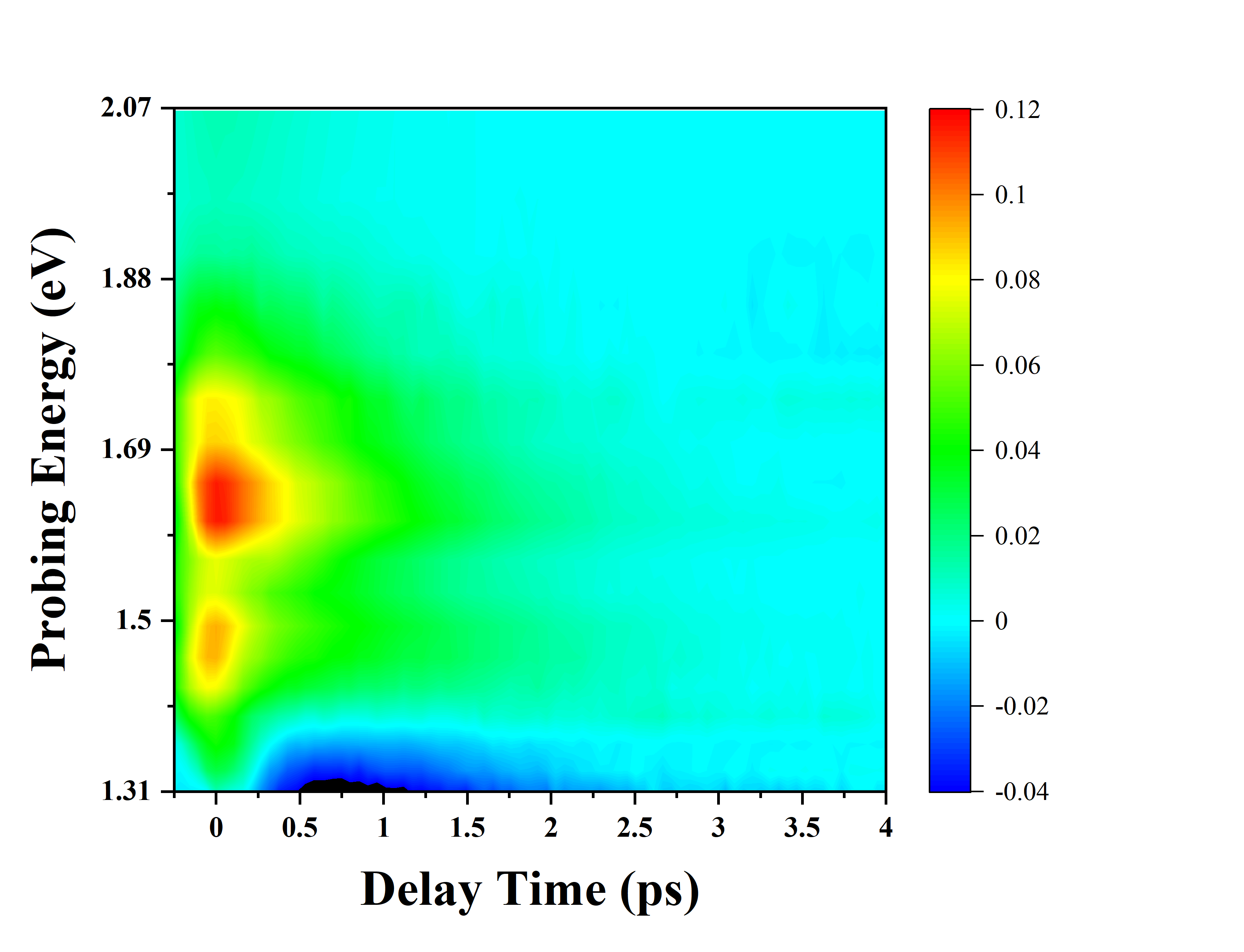}
\caption{\textbf{Time--resolved transient transmittance spectra of InN.} Transient transmittance spectra measured at a pump fluence of 63.8 J/m$^2$, using a pump laser with photon energy of 1.55 eV. The probe photon energies range from 1.31 to 2.07 eV.}
\label{contour}
\end{center}
\end{figure}

\begin{figure*}[ht]
  \begin{minipage}[b]{0.33\textwidth}
    \includegraphics[width=\linewidth]{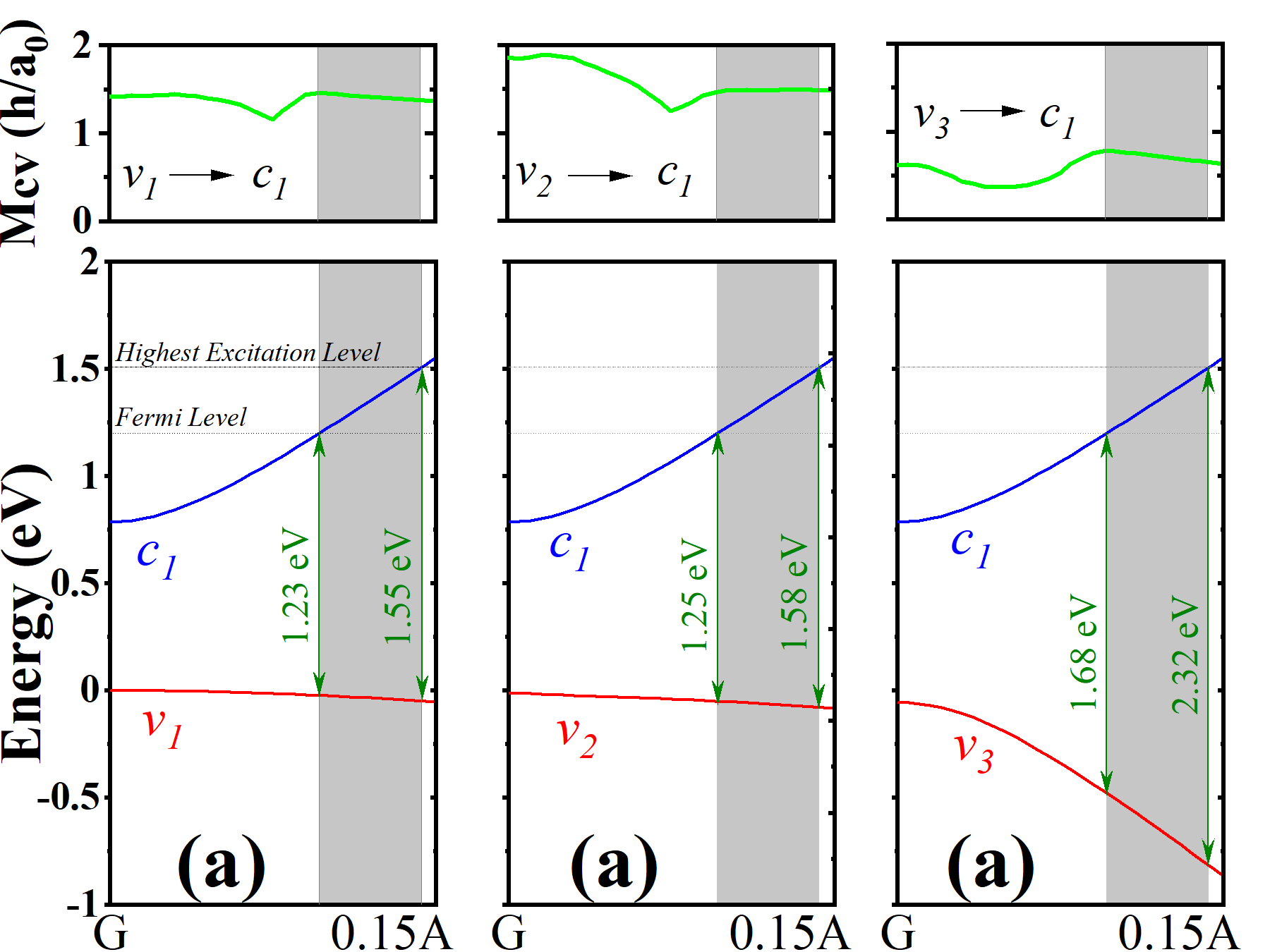}
  \end{minipage}%
  \begin{minipage}[b]{0.33\textwidth}
    \includegraphics[width=\linewidth]{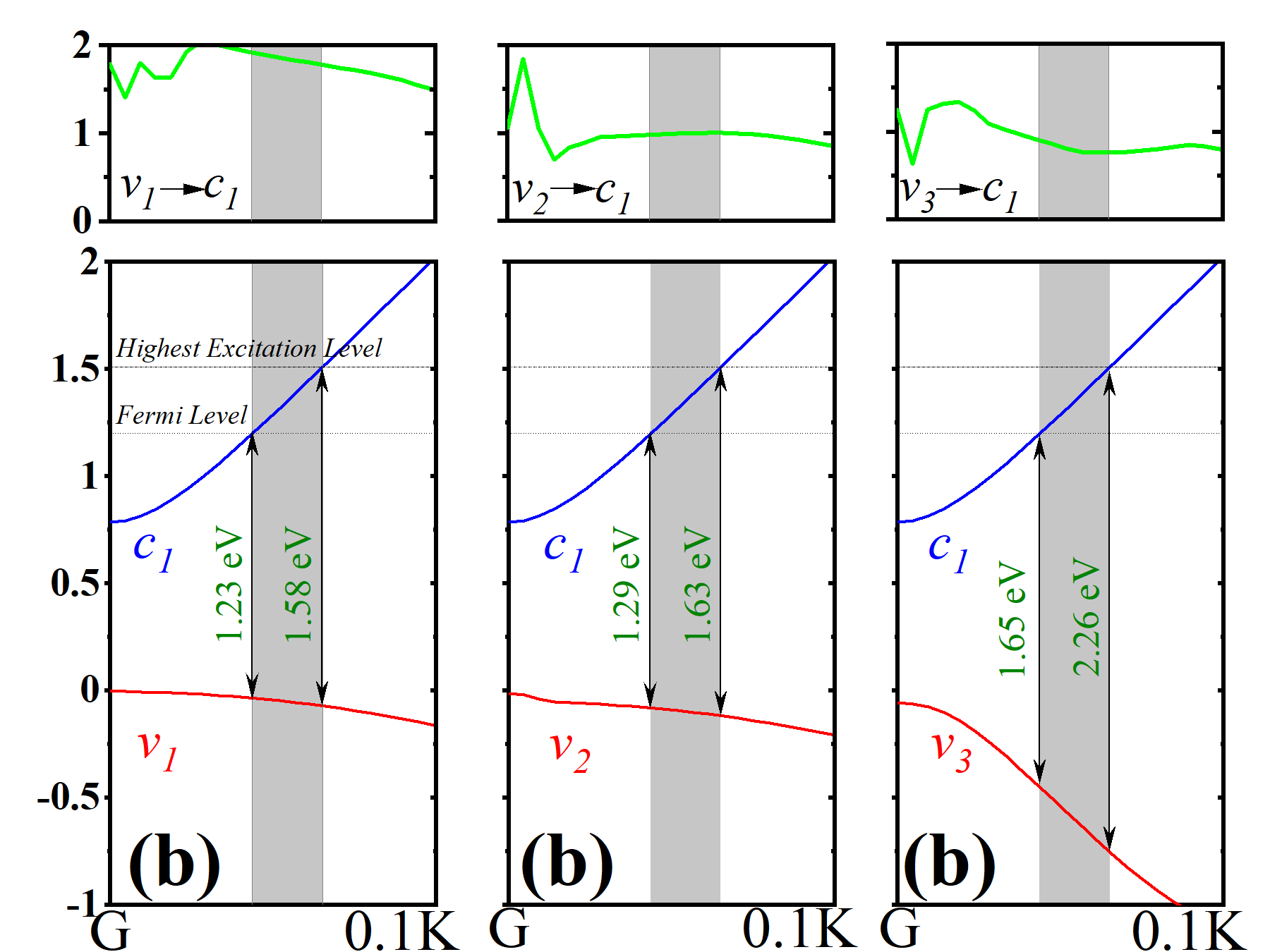}
  \end{minipage}%
  \begin{minipage}[b]{0.33\textwidth}
    \includegraphics[width=\linewidth]{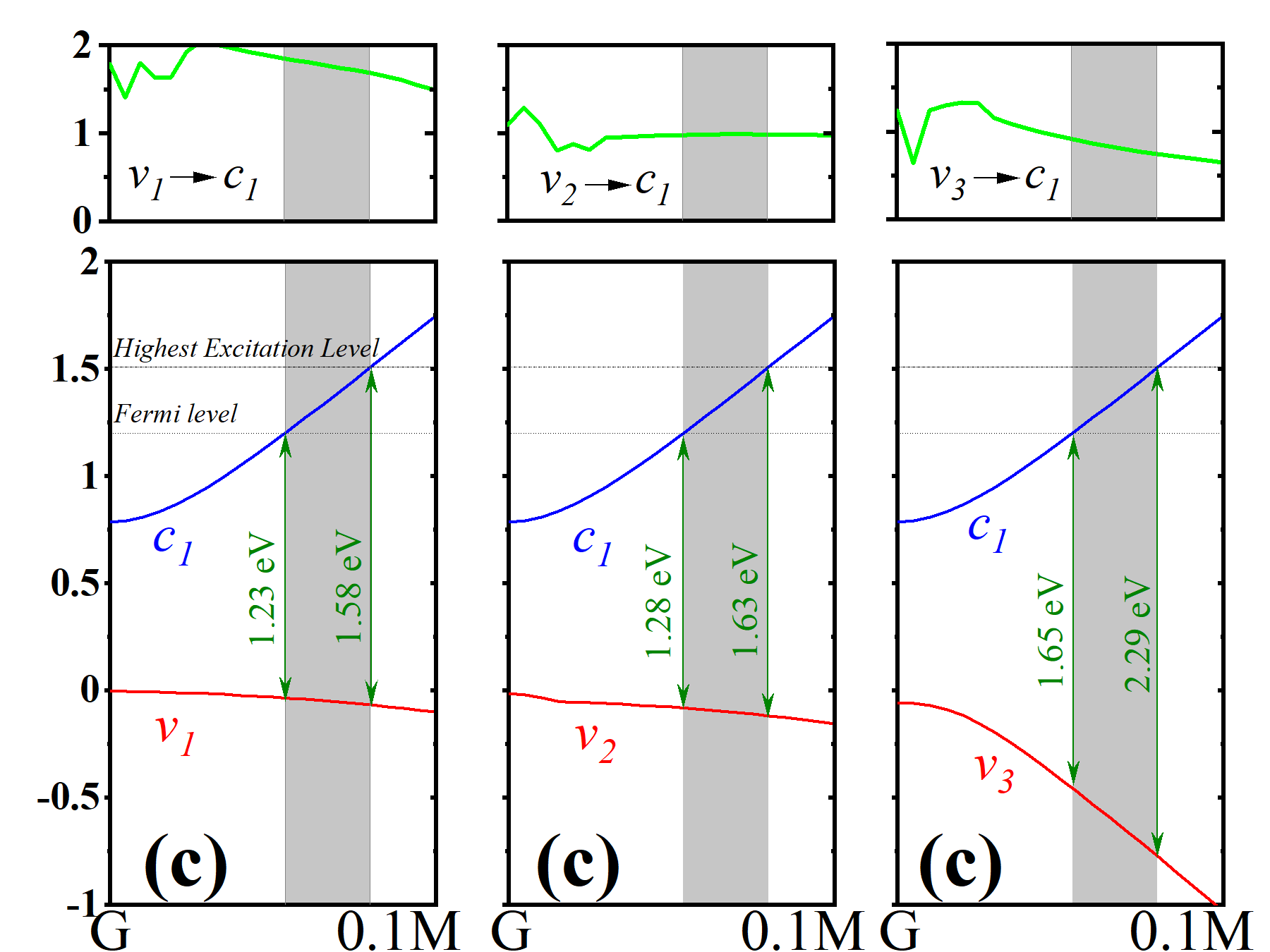}
  \end{minipage}
\caption{\textbf{Band structures and optical transition matrix elements of wurtzite InN.} Band structures illustrating possible optical transitions from the valence bands ($v_1$, $v_2$, and $v_3$) to the lowest conduction band $c_1$ along three high--symmetrical paths ((a): $\Gamma\rightarrow$ A, (b): $\Gamma\rightarrow$ K, and (c): $\Gamma\rightarrow$ M). The upper panels show the corresponding optical matrix elements, which are proportional to the transition probability. The dash--dotted lines denote the highest excitation level provided by a 1.55 eV pump laser, while the dotted lines indicate the Fermi level at equilibrium before photoexcitation. Shaded gray regions represent the energy window accessible to the probe light.}  
\label{OpticalTransition}
\end{figure*}

\textbf{\textcolor{black}{Optical Switching}}: Figure \ref{contour} shows that transient transmittance spectrum measured across the probe energy range of 1.31--2.07 eV under optical excitation at 1.55 eV with a pump fluence of 63.8 J/m$^2$. The transient transmittance signal is defined as $dT/T$=[$T(t)$-$T(0)$]/$T(0)$, where $T(t)$ and $T(0)$ represent the time--dependent and static transmittance, respectively. The overall temporal response shows two notable switching regions, suggesting the potential of InN as a multicolored optical switching material, where ``multicolor" refers to distinct optical response at multiple probe photon energies. The most pronounced switching region is centered at approximately 1.63 eV, while another switching region is located at around 1.43 eV. These switching states are attributed to \textit{transient Pauli blocking effect}, which induces a reversible switching from an opaque to a relatively transparent state \cite{Jia2025}. The underlying mechanism can be interpreted by considering the allowed optical transition pathways from different valence bands to the lowest--lying conduction band ($c_1$), as shown in Fig. \ref{HSE}(a).

From the perspective of the pump--probing measurement principle, when electrons are excited and injected into the conduction band by the pump laser, the probe laser can detect their transient occupancy. Based on the calculated band structure, we evaluated the optical transition probability from the valence bands to the conduction band, which help to explain the optical switching behavior observed in Fig. \ref{contour}. The possible interband transitions near the $\Gamma$ point are illustrated in the upper panels of Figs. \ref{OpticalTransition}, where optical transitions from $v_1\rightarrow c_1$, $v_2\rightarrow c_1$, and $v_3\rightarrow c_1$ are all dipole--allowed. Taking into account the highest excitation level provided by the 1.55 eV pump laser and the estimated static Fermi level at 1.20 eV, a broad transient transmission window ranging from 1.23 to 2.32 eV is expected, as shown in Fig. \ref{OpticalTransition}. This range is in rough agreement with the experimentally observed \textit{dT/T} spectrum from 1.31 to 2.07 eV in Fig. \ref{contour}. A more quantitative interpretation, however, requires incorporating the thermal broadening of the Fermi–Dirac distribution at elevated electron temperature, which will be discussed in the final section.

\begin{figure}[hb]
\begin{center}
\includegraphics[clip, width=9cm]{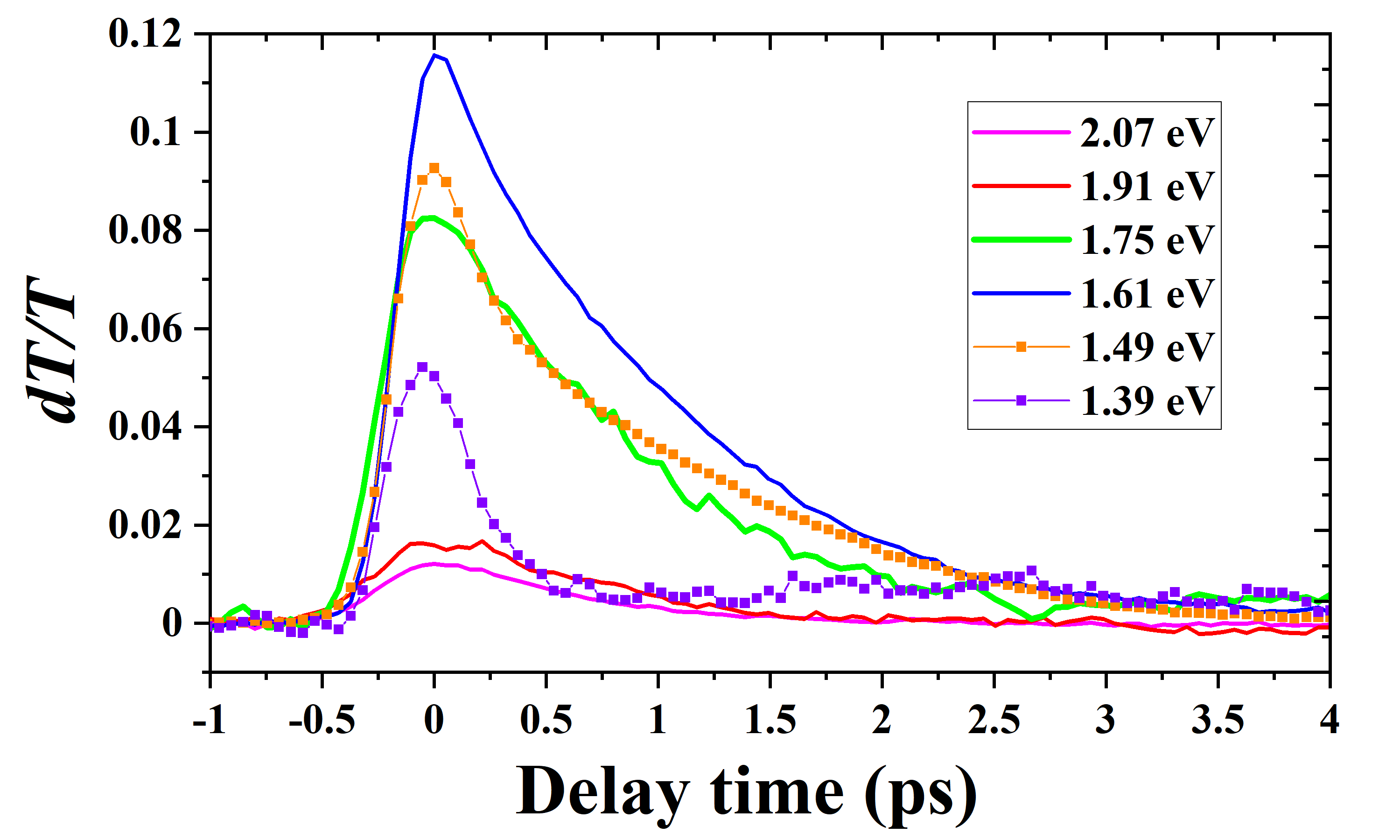}
\caption{\textbf{Energy--resolved transient transmittance dynamics in InN.} Transient $dT/T$ traces measured under 1.55 eV excitation at a pump fluence of 63.8 J/m$^2$, with the probing energy systematically varied to capture energy--dependent carrier dynamics.}
\label{trace}
\end{center}
\end{figure}

The emergence of two distinct optical switching centers can be attributed to the presence of multiple allowed optical transition pathways between the valence and conduction bands. For a given photon energy, the final conduction--band state may be accessed via more than one interband transition channel. The switching feature centered at 1.43 eV is primarily associated with optical transitions from the $v_1\rightarrow c_1$ and $v_2\rightarrow c_1$ bands, both of which exhibit high optical transition probability ($M_{cv}$), as shown in Fig. \ref{OpticalTransition}. The higher--energy switching feature near 1.63 eV is attributed to optical transitions from the $v_2\rightarrow c_1$ and $v_3\rightarrow c_1$ bands. Notably, transient transmission persists even above 1.70 eV in Fig. \ref{contour}, exceeding the pump photon energy of 1.55 eV, which is attributed to the $v_3\rightarrow c_1$ transition. These observations confirm that the \textit{transient Pauli blocking effect} enables broadband optical switching, spanning a spectral range wider than the excitation energy.    

It is also worth noting that a negative signal appears around 1.31 eV for delays starting from approximately 0.25 ps. This feature is attributed to absorption arising from optical transitions between the valence band and the conduction band states near Fermi level. After photoexcitation, some electrons close to the Fermi level are thermalized to higher energies through scattering processes, leaving behind empty states (“hole”) close to the Fermi level. These empty states allow additional optical transitions of photoexcited electrons, i.e., absorption, resulting in the negative feature observed in the $dT/T$ traces.


\textbf{\textcolor{black}{Transient Transmittance Trace}}: Fig. \ref{trace} presents the $dT/T$ traces measured at different probe photon energies. The dependence of the peak $dT/T$ intensity on probe energy is consistent with the contour map shown in Fig. \ref{contour}. All traces exhibit a rapid rise to a maximum followed by a subsequent decay. The initial rise is commonly attributed to the rapid population of the conduction band states by thermalized electrons via electron--electron scattering processes \cite{Crepaldi2012}, after which the thermalized electrons gradually recombine with holes in the valence band. However, the total photoexcited electron density in our experiment is estimated to be only 3.84 $\times$ 10$^{16}$ cm$^{-3}$, which is insufficient to directly induce a broadband Pauli blocking effect. We therefore hypothesize that the energy of the photoexcited electrons transferred to the conduction electrons through energy and momentum relaxation, forming a ``hot" thermalized FD distribution. Under this assumption, the observed $dT/T$ signal at a given probe energy reflects the temporal evolution of the electronic occupation, namely the cooling dynamics of the ``hot" thermalized FD distribution. 

For an isotropic parabolic band dispersion, a given transition energy (\textit{e.g.}, the probe photon energy 1.55 eV) corresponds to a single or only a few $k$--points in $k$--space. When the bands are anisotropic, however, the situation becomes more complex. For instance, in diamond structured Ge, the energy separation between the valence and conduction bands remains nearly constant over a range of $k$--values along the (111) direction, leading to an extreme case where the interband critical point is two-dimensional in character \cite{Xu2017}. In contrast, for wurtzite InN, the effective electron mass parallel ($m_{||}$=0.068$m_0$) and perpendicular ($m_\perp$=0.070$m_0$) to the $c$ direction \cite{Gorczyca2008} are nearly identical, indicating only a modest anisotropy in its electronic band structure. Therefore, a given transition energy in InN is expected to involve electronic states at a single (or a few) $k$--points, rather than an extended $k$--space region. 

On the other hand, at a fixed photon energy, the final conduction--band states involved in the optical transition can still couple to multiple valence--band states, as illustrated in Fig.~\ref{OpticalTransition}. For instance, despite the relatively small energy separation between the $v_1$ and $v_2$ valence bands, a given probe photon energy may excite transitions from both $v_1\rightarrow c_1$ and $v_2\rightarrow c_1$, leading to the detection of different conduction band states above the static Fermi level, as illustrated in Fig. \ref{OpticalTransition}. Consequently, the measured $dT/T$ traces may reflect carrier thermalization dynamics across multiple energy levels. This effect is particularly evident for probe energies below 1.49 eV, where both $v_1\rightarrow c_1$ and $v_2\rightarrow c_1$ transitions are allowed and the transient traces exhibit pronounced bi-- or multi--exponential decay features, as shown in Fig. \ref{trace}. To avoid overinterpretation of the $dT/T$ traces that may involve multiple optical transitions, we focused on those measured at probe energies above 1.61 eV, where the signal predominantly originates from the $v_3\rightarrow c_1$ transition. Although slight variations still exist among different $k$--space directions (\textit{e.g.}, $\Gamma\rightarrow A$, $\Gamma\rightarrow K$, and $\Gamma\rightarrow M$), the corresponding transitions at a given probe energy can be reasonably approximated by a single dominant channel owing to their small energy separations. 




To extract the characteristic cooling time of ``hot" thermalized FD distribution, the $dT/T$ traces were first fitted using a simple exponential relaxation function: $dN/dt=G(t) - N/\tau$, where $N$ is the number of thermalized electron contributing to the signal at the probe level, and $\tau$ represents the characteristic cooling time. The generation term, $G(t)\propto\exp\left[-2\left( t/t_p\right)^2\right]$, describes carrier injection induced by a Gaussian pump pulse with the pulse width $t_p$. For probe photon energies below 1.49 eV in Fig. \ref{trace}, the $dT/T$ traces exhibit bi-- or multi--exponential characteristics, likely due to overlapping contributions from multiple optical transitions. In contrast, for probe photon energies above 1.75 eV in Fig. \ref{trace} , a single exponential fit adequately reproduces the measured dynamics. Notably, as the probe energy increases from 1.75 to 2.07 eV, the extracted $\tau$ gradually decreases from 0.90 ps to 0.66 ps, which are slightly smaller than the previously reported characteristic times \cite{Chen2003, Su2010, Wen2006}. This trend indicates that higher probe photon energies correspond to higher--energy states within the hot thermalized FD distribution, where carrier cooling proceeds more rapidly. Such an energy dependence suggests that the cooling dynamics are primary governed by energy relaxation through electron–-phonon coupling, and that a more accurate description, accounting for the energy--dependent nature of the electron distribution, is required to model the cooling dynamics of the thermalized FD distribution.

To further elucidate the cooling mechanism, we quantitatively evaluated the influence of defect--induced momentum scattering using the measured mobility. The InN film used in this study exhibited a mobility of $\mu$=5.43 cm$^2$/Vs. The average scattering time $\tau_s$ during transport can be estimated from $\mu$=$e\tau_s/m^*_e$, where $e$ is the elementary charge. Taking $m^*_e$= 0.13$m_0$, the calculated average scattering time is approximately 0.40 fs. This value is much shorter than the typical thermalization time on the order of picoseconds, indicating that momentum randomization caused by defect--related scattering, such as ionized impurity scattering \cite{Jia2014}, occurs much faster than energy redistribution. Consequently, such scattering mainly affects momentum relaxation but has only a limited impact on the overall carrier thermalization dynamics. 

\textbf{\textcolor{black}{Modeling the cooling of the FD distribution}}: The cooling process of a hot thermalized FD distribution is primarily governed by electron--phonon interactions, which can be effectively described by the classical two--temperature model. Assuming that the electron--electron and phonon--phonon interactions are sufficiently fast to maintain local equilibrium within the carrier and phonon populations, the temporal evolution of the electron temperature $T_e$ and the lattice temperature $T_l$ can be described by evaluating the electron energy loss rate to the lattice using the Boltzmann equation \cite{Fatti2000}, resulting in the two--temperature model \cite{Sun1993, Kaganov1957}

\begin{equation}
\begin{cases}
C_e(T_e) \dfrac{\partial T_e}{\partial t} = I(t) - g_{ep} (T_e - T_l), \\[6pt]
C_l \dfrac{\partial T_l}{\partial t} = g_{ep} (T_e - T_l),
\end{cases}
\label{TTM}
\end{equation}
where $C_e$ ($C_l$) is the heat capacity of electrons (lattice), $g_{ep}$ is the electron--phonon coupling coefficient. For InN, its heat capacity is $C_{\rm InN}$=2.197$\times$10$^6$ Jm$^{-3}$K$^{-1}$ \cite{Barin1977}. For electronic temperatures below a few thousand Kelvin, a linear dependence of the electronic heat capacity, $C_e(T_e)$=$\gamma T_e$, has been found to be a good estimation \cite{Singleton}. The temporal evolution of laser intensity is given as \cite{Alexopoulou2024, CD2023}

\begin{equation}
I(t)=\sqrt{\frac{\beta}{\pi}}\frac{AI_0}{t_p}\exp\left[-\beta\left( \frac{t-2t_p}{t_p}\right)^2\right],
\label{laser}
\end{equation}
where $\beta=4ln(2)$, $I_0$ is the pumping fluence, $A=(1-R)(1-e^{-\alpha d})$ is the effective absorption coefficient. Note that we neglect thermal diffusion processes in Eqs. (\ref{TTM}). 

The cooling of the electron temperature leads to a gradual narrowing of the FD distribution from its thermally broadened profile. From the aspect of the pump--probe measurement, the temporal evolution of occupation probability of electrons at a given energy level ($E_p$) in the conduction band for degenerate semiconductors can be described by the following FD distribution
\begin{equation}
f(E_p) = \frac{1}{\exp(\frac{E_p-E_f(t)}{k_BT_e(t)})+1}, 
\label{FD}
\end{equation}
where $k_B$ is the Boltzmann constant and $E_f(t)$ denotes the time--dependent Fermi level, which evolves due to changes in the electronic temperature and the photoexcited carrier density in the conduction band. The electron population at $E_p$ can then be expressed as 

\begin{equation}
n_p(t)=\int_{E_p}^{E_p+\delta E} f(E_p, E_f(t), T_e(t))\cdot D(E_p) dE, 
\label{np}
\end{equation}
where $\delta E$ represents a small energy interval around $E_p$, corresponding to the energy width of the probe laser, and $D$($E_p$) is the density of states at energy $E_p$. Taking into account the nonparabolic nature of conduction band in InN, the $E$--$k$ dispersion can be approximated by retaining the second--order $k^2$ ellipsoidal energy surfaces and introducing a nonlinear dependence on the energy, namely, 
\begin{equation}
\frac{\hbar^2k^2}{2m_0^*} =\gamma(E)=E+\alpha E^2 + \beta E^3+ \cdots,
\label{nonlinear}
\end{equation}
where $m_0^*$ is the electron effective mass at the conduction band minimum, $\alpha$ and $\beta$ are the nonparabolicity coefficients of the conduction band, $E$ is the electron energy relative to the conduction band minimum (CBM). By keeping only the first nonlinear term, the well--known Kane quasilinear dispersion relation is obtained \cite{Kane1957}: 
\begin{equation}
\frac{\hbar^2k^2}{2m_0^*} =E(1+\alpha E).
\label{kane}
\end{equation}
From Eq. (\ref{kane}), the $E$--$k$ dispersion can be obtained, and then the nonparabolic $D$($E$) can be derived as follow:
\begin{equation}
D(E) = \frac{1}{2\pi^2} \left( \frac{2m_0^*}{\hbar^2} \right)^{3/2} \sqrt{E(1 + \alpha E)} \cdot (1 + 2\alpha E).
\label{DOS}
\end{equation}
Our simulation shows that $\alpha$=0.20 eV$^{-1}$ accurately reproduces the Fermi level shift in the InN film under investigation.  

Eq. (\ref{np}) describes the time dependence of $n_p(t)$ at the probing level, which results from the evolution of the electron temperature $T_e(t)$ and $E_f(t)$ during the cooling process. From the aspect of the Pauli blocking effect, this quantity is considered proportional to the measured $dT/T$ intensity. However, solving Eq. (\ref{np}) requires accurately determining $E_f(t)$, which depends on both the electron temperature and the electron concentration in the conduction band. For a known electron density $n_{total}(t)$ at time $t$, $E_f(t)$ can be determined using a numerical bisection root-finding approach applied to the following equation

\begin{equation}
n_{total}(t)=\int_{E_c}^{\infty} f(E, E_f(t), T_e(t))\cdot D(E) dE,
\label{ntotal}
\end{equation}
where $n_{total}(t)=n_0+n_{ex}(t)$. Because $n_{ex}$= 3.84 $\times$ 10$^{16}$ cm$^{-3}$ is three orders of magnitude smaller than $n_0$, its contribution to $E_f(t)$ is negligible. Therefore, we assume $n_{total}(t)=n_0$. Under this condition, the Fermi level $E_f(t)$ can be determined self--consistently from the electron temperature $T_e(t)$ through the carrier--density conservation.

\begin{figure}[ht]
\begin{center}
\includegraphics[clip, width=9cm]{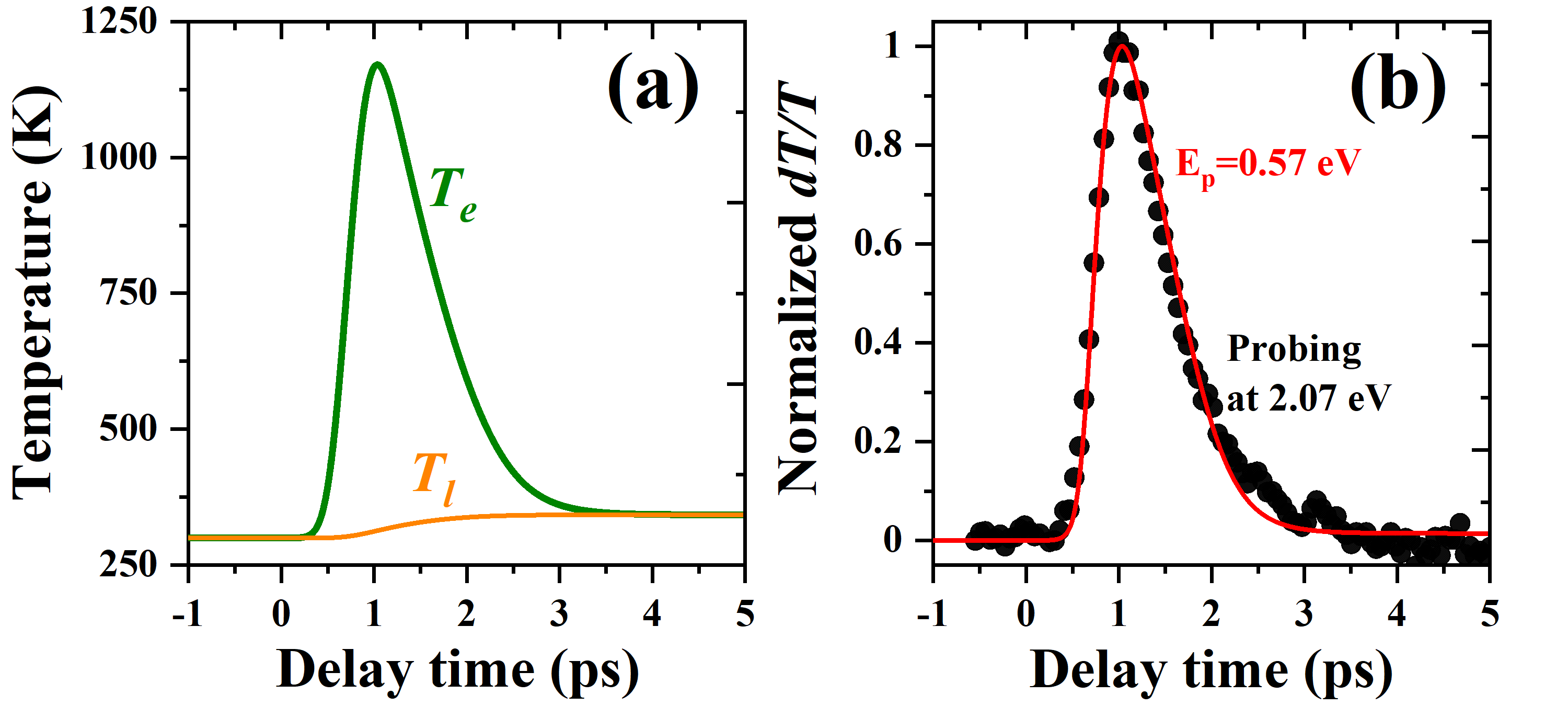}
\caption{\textbf{Comparison between calculated and measured carrier dynamics.} (a) Calculated temporal evolution of electron ($T_e$) and lattice ($T_l$) temperature, using $g_{ep}$ = 1.0 $\times$ 10$^{17}$ W/m$^3$$\cdot$K and $\gamma$ = 1.52 mJ/mol$\cdot$K$^2$. (b) Experimental $dT/T$ data (circular markers) obtained under 1.55 eV excitation with a pump fluence of 63.8 J/m$^2$, probed at 2.07 eV. The solid line represents the calculated temporal evolution of electron occupation at $E_p$=0.57 eV above the CBM. }
\label{FDsimulation}
\end{center}
\end{figure}

Following the above model, the temporal evolution of $n_p(t)$ at the probing level $E_p$ can be obtained by first calculating $T_e(t)$ in Eq. (\ref{TTM}) with an assumed set of parameters ($g_{ep}$ and $\gamma$) using the fourth order Runge--Kutta method. Then, $E_f(t)$ is determined self--consistently from $T_e(t)$ using Eq. (\ref{ntotal}). Finally, $n_p(t)$ is calculated based on Eq. (\ref{np}). Since $n_p(t)$ directly reflects the experimentally measured $dT/T$ traces, we compared the calculated $n_p(t)$ with the measured $dT/T$ traces, which are mainly governed by approximately single--pathway optical processes corresponding to the $v_3 \rightarrow c_1$ transition. We found that using an approximately constant $g_{ep}$=1.0 $\times$ 10$^{17}$ W/m$^3$$\cdot$K and an electronic heat capacity coefficient $\gamma$ in the range of 1.52 to 2.02 mJ/mol$\cdot$K$^2$ can qualitatively reproduce the shape of the experimental $dT/T$ traces. The best agreement is then achieved by tuning $E_p$ with a slight adjustment of $\gamma$. The obtained $E_p$ values systematically increases with increasing the probing energy. Note that in Eq. (\ref{FD}), $E_p$ is defined as the energy separation between the probed final state in the conduction band and the CBM, whereas the probing energy in the experiments corresponds to the optical transition from the valence band to the conduction band. For the $v_3 \rightarrow c_1$ transition in InN, the $c_1$ band energy increases with the wavevector $k$, while the $v_3$ energy decreases. Therefore, a higher probing energy corresponds to a higher final state energy above the CBM, i.e., a larger $E_p$.

Fig. \ref{FDsimulation} presents a representative comparison between calculation and experiment. Fig. \ref{FDsimulation}(a) shows the calculated temporal evolution of the electron and lattice temperatures, $T_e$ and $T_l$, at $E_p$=0.57 eV above the conduction band minimum, using $g_{ep}$ =  1.0 $\times$ 10$^{17}$ W/m$^3$$\cdot$K and $\gamma$ = 1.52 mJ/mol$\cdot$K$^2$. Fig. \ref{FDsimulation}(b) compared the experimentally measured traces probed at 2.07 eV with the calculated $dT/T$ trace at $E_p$=0.57 eV above the conduction--band minimum. The close agreement between them confirms that the model successfully captures the underlying dynamics of the $dT/T$ response. It should be mentioned that both $g_{ep}$ and $\gamma$ were treated as effective fitting parameters within the two-temperature model. An approximately constant $g_{ep}$ successfully reproduces all measured transient transmittance traces. Given the lack of reported $g_{ep}$ values for InN, we adopted 1.0 $\times$ 10$^{17}$ W/m$^3$$\cdot$K, which lies within the range commonly reported for metals such as aluminum (2.45 $\times$ 10$^{17}$ W/m$^3$$\cdot$K) \cite{Lin2008} and copper (1.00 $\times$ 10$^{17}$ W/m$^3$$\cdot$K) \cite{Elsayed-Ali1987}. Moreover, a $\gamma$ value in the range of 1.52 to 2.02 mJ/mol$\cdot$K$^2$ yields excellent agreement with the experimental transient data under femtosecond laser excitation. This value is comparable to those measured in metals such as Al or In \cite{Kittel}. The theoretical value, estimated from 
\begin{equation}
\gamma=\frac{\pi^2 k_B^2}{3}D(E_F) 
\end{equation}
where $D(E_F)$ denotes the density of state at the Fermi level (0.40 eV) in equilibrium, is only 0.0064 mJ/mol$\cdot$K$^2$, reflecting the intrinsically low density of states in InN even when the Fermi level is relatively high. Such a large discrepancy should be understood in terms of the fundamentally different origin of the electronic excitation. In our case, the energy is stored in the hot FD distribution generated by femtosecond photoexcitation, where a significant portion resides in the high--energy tail above the Fermi level. This contrasts sharply with thermal excitation under equilibrium conditions, in which the carrier energy is primarily concentrated near the Fermi level. Therefore, the experimentally extracted $\gamma$ represents a non-equilibrium contribution of photoexcited carriers and should not be directly compared with values derived from static low--temperature measurements.

Finally, we highlight several key implications of our findings. Our results demonstrate that the \textit{transient Pauli blocking effect} can be effectively induced by a laser--driven rise in electronic temperature as shown in Fig. \ref{FDsimulation}(a), rather than by the injection of a large number of electrons into the conduction band \cite{Jia2025}. Our experiments further indicate that directly extracting thermalization or cooling times by fitting the transient transmittance traces with single-- or multi--exponential functions may lead to misinterpretation, and that the underlying physical mechanisms must be carefully examined. Importantly, the theoretical framework based on a quasi–-equilibrium Fermi–-Dirac distribution proposed here provides a unified description of optical switching arising from \textit{transient Pauli blocking effect}, and enables direct evaluation of the corresponding spectral switching range. These fundamental insights offer practical guidance for the laser--driven design of ultrafast optical switching devices, paving the way for next-generation photonic technologies such as all--optical communication and quantum information processing.

\section{Conclusions}

In this study, we explore ultrafast optical switching in InN thin films using pump--probe transient transmittance measurement, combined with first--principles electronic band structure calculations. By resolving the transient optical response across a range of probe photon energies, we disentangle the contributions of multiple interband transition pathways from the valence bands to the lowest lying conduction band. Moreover, our results reveal that conventional exponential fitting models often misinterpresent the electron cooling dynamics, primarily due to the different cooling time at different probing level.

To gain deeper physical insights, we establish a theoretical framework based on a quasi--equilibrium FD distribution, which enables quantitative reproduction of the transient transmittance traces. We show that the observed ultrafast dynamics can be accurately described by a thermalized FD distribution with an elevated temperature, allowing precise prediction of transient optical transparency windows. Based on the experimental results, we further extract an electron-–phonon coupling constant of $g_{ep}$=1.0 $\times$ 10$^{17}$ W/m$^3$$\cdot$K, and an electronic specific heat coefficient in the range of 1.52 to 2.02 mJ/mol$\cdot$K$^2$ for InN. 

Crucially, we demonstrate that the \textit{transient Pauli blocking effect} can arise from a temperature--induced redistribution of electrons within the conduction band due to laser excitation, rather than from the injection of a large number of electrons. These findings provide a refined understanding of non--equilibrium carrier dynamics in semiconductors and opens new avenues for laser--driven material design of ultrafast optical switching components, including modulators, filters, and shutters for next--generation photonic applications such as all--optical networking or optical computing.


\section{Acknowledgments}
J. Jia acknowledges the funding support from JSPS KAKENHI Grant--in--Aid for Scientific Research (B) (Grant No. 25K01862) and from Waseda University Grant for Special Research Projects (Project No. 2025C–-157). R. Nakazawa, K. Fukutani, and S. Kera acknowledges support from JSPS KAKENHI Grant--in--Aid for JSPS Fellows (Grant No. 25KJ0391).

\section{Author contributions}
JJ initiated and formulated the research idea. JJ and TM fabricated the InN film. MK and YS conducted transmittance and reflectance experiments. JJ and TY developed the pump--probe measurement system and collected the experimental data. RN, KF and SK carried out the HS--UPS measurements. JJ wrote/revised the manuscript. All authors reviewed the manuscript.

\section{Data Availability}
The data that support the findings of this study are available from the corresponding author, JJ, upon reasonable request.

\section{Competing interests}
The authors declare no competing interests.
\end{document}